\documentclass[journal,10pt, twoAt t thcolumn]{IEEEtran}

 \usepackage[T1]{fontenc} 
\usepackage{url}
\usepackage{amsmath}
\usepackage{subcaption}
\usepackage{cite}
\usepackage{color}
\usepackage{graphicx}

\usepackage{amsmath}
\usepackage[cmintegrals]{newtxmath}
\usepackage{fixltx2e}
\usepackage{stfloats}
\usepackage{array}
\usepackage{booktabs}
\usepackage{multirow}
\usepackage[font={small}]{caption}

\begin{document}
\title{Communication Through Breath: Aerosol Transmission}
\author{{ Maryam~Khalid, Osama~Amin, Sajid~Ahmed, Basem~Shihada and Mohamed-Slim~Alouini}

\thanks{M. Khalid is with Department of Electrical and Computer Engineering, Rice University, Houston, TX 77005 USA. E-mail : maryam.khalid@rice.edu.

O. Amin, B. Shihada and  M.S. Alouini are with CEMSE Division,  King Abdullah University of Science and Technology (KAUST),  Thuwal, Makkah Province, Saudi Arabia. E-mail: \{osama.amin, basem.shihada, slim.alouini\}@kaust.edu.sa.

S. Ahmed is with Electrical Engineering Department, Information Technology University, Lahore 54000, Pakistan. Email:  sajid.ahmed@itu.edu.pk. 
}} 

\maketitle

\begin{abstract}
  Exhaled breath can be used in retrieving information and creating innovative communication systems. It contains several volatile organic compounds (VOCs) and biological entities that can act as health biomarkers. For instance, the breath of infected human contains a nonnegligible amount of pathogenic aerosol that can spread or remain suspended in the atmosphere. Therefore, the exhaled breath can be exploited as a source's message in a communication setup to remotely scan the bio-information via an aerosol transmission channel. An overview of the basic configuration is presented along with a description of system components with a particular emphasis on channel modeling. Furthermore, the challenges that arise in theoretical analysis and system development are highlighted. Finally, several open issues are discussed to concretize the proposed communication concept.
\end{abstract}


\IEEEpeerreviewmaketitle

\section{Introduction}\label{sec:intro}

Information and communication technology (ICT) has inspired the development of numerous non-conventional applications and broadened the scope of communication systems to versatile horizons.  Recently, HBC has been introduced under the ICT umbrella to create comprehensive access to the five human senses through various communication technologies \cite{dixit2017human}.  The objective of this article is to add an extra member to the existing pool of human body's features that can be tapped as a potential information reservoir. The exhaled breath contains biomarkers that can be used in several bio-inspired applications such as monitoring health-care and detecting diseases. Therefore, we propose, for the first time, a holistic communication system that exploits the human breath as a \textit{source of information}.

Human breathing is a process that involves the interaction of internal organs (Lungs) with the atmosphere. Thus, it is very likely that the exhaled breath contains \textit{footprints} from inside the body. It can contain not only particles from the respiratory tract but also blood-borne compounds that enter the exhaled breath during exchange taking place in the alveoli \cite{almstrand2011analysis, de2014review}. A recent review study identified 872 VOCs in the respiration of health humans \cite{de2014review}. Interestingly, the VOCs are shaped according to the human health, age, diet, sex, body fat, height, behavioral/lifestyle differences, and other biological characteristics \cite{de2014review}. Several efforts in medicine and clinical research made use of the breath biomarkers and developed methods for sampling and analyzing the exhaled breath \cite{flu1, foot, voc}.
For instance, the presence of viruses such as human influenza A \cite{flu1} and foot-and-mouth disease \cite{foot} in exhaled breath has been confirmed through different experiments conducted on the exhaled breath. Moreover, the Breast/Lung cancer and other diseases can be diagnosed by detecting volatile organic compounds in exhaled air \cite{voc}.

\begin{figure}
	\centering
\includegraphics[width=3.5in]{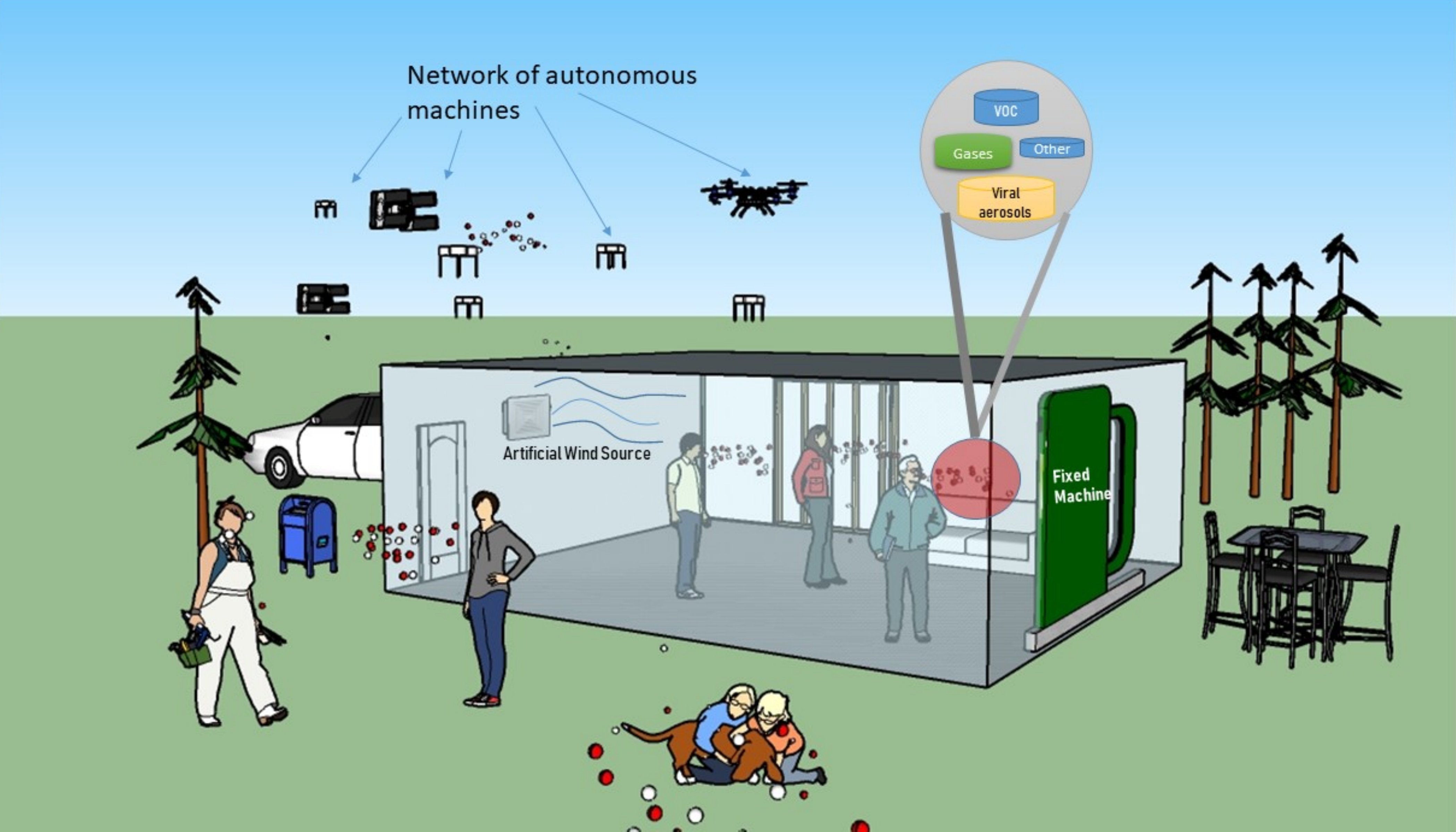}
\caption{Breath communication systems overview. }
\label{overview}
\end{figure}

In this article, we exploit the inhaled and exhaled breath in a novel way that would lead to versatile macro-scale applications. Instead of sampling and analyzing the exhaled air on lab-scale environments, we propose considering it as a component of a communication system where the scope expands to the complete process of information exchange as depicted in Fig.~\ref{overview}. As a candidate communication technology, MC is considered due to its compatibility with biological entities, where molecules or chemical signals are deployed as information carriers \cite{farsad2016comprehensive}. Thus, adopting breath as an information source in an MC setup introduces a new dimension to the areas of chemical-signaling communication and~HBC.

In the rest of this article, we focus the discussion on the potential and feasibility of breath as an information source and introduce it in the context of an MC system in a generic manner. To establish a detailed picture of our proposed idea, we explore aerosol transmission in depth. In particular, we consider pathogenic aerosols that are virus-laden micro-sized droplets which remain suspended in air and are responsible for diseases spread as shown in Fig. \ref{overview}.  Sneezing, coughing, talking and breathing are some of the originating sources for these aerosols. For ease of understanding, we gather these four mechanisms under one umbrella and refer to them as \textit{exhaled breath}. It is worth to mention that the same concept can be used in analyzing the communication scenario through inhalation while taking into consideration the effect of human body interaction. In Section \ref{sec:block}, we provide an overview of the breath-based communication system, followed by a detailed discussion on channel modeling in Section \ref{sec:channel}. Then, a case study on Gaussian dispersion models is discussed while highlighting the possible challenges in Section \ref{sec:case}. Several open research issues are discussed in Section \ref{sec:issues} followed by the conclusion in Section \ref{sec:conc}.

\section{Breath Communication System : System Overview} \label{sec:block}
\begin{figure}
\centering 
\includegraphics[width=3.5in]{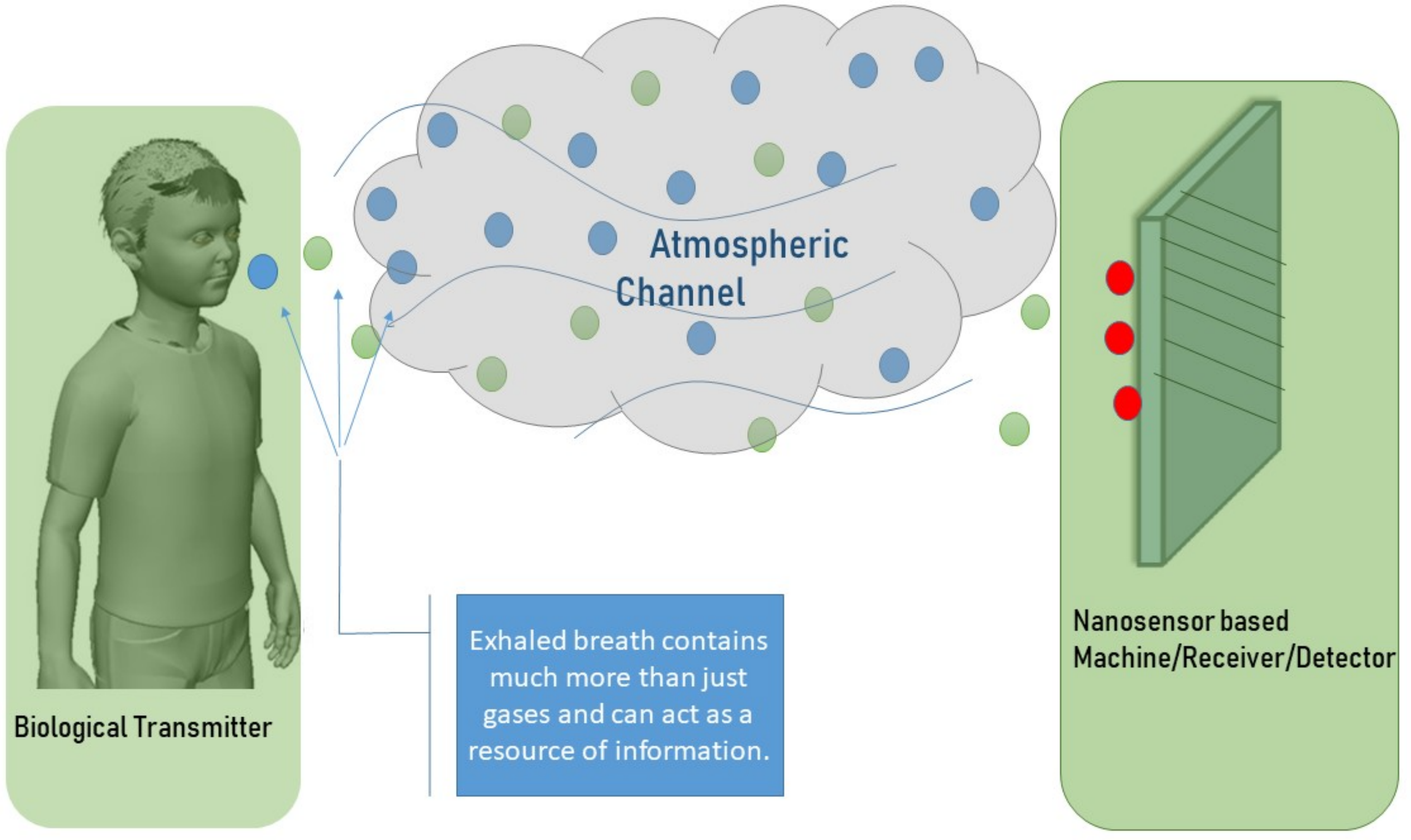}
\caption{Breath communication system components.}
\label{block}
\end{figure}

In this section, we present the basic components of a communication system where the aerosol acts as the information carrier. Similar to traditional communication systems, the setup is composed of three blocks, biological transmitter, biological receiver, and transmission channel as shown in Fig.~\ref{block}.

\subsection{Biological Transmitter}

The focus of this article is on information extraction from exhaled breath, where the human plays the role of a \textit{transmitter}. Throughout the breath communication system, different types of messages can be transmitted and retrieved to reflect human health information and characterize several biological features \cite{de2014review}. Two critical factors at the transmitter side can affect the communication system design and performance. The first factor is the transmitter mobility status, where having a stationary emission source simplifies the analysis and design \cite{aerosol2018}.  However, having mobile transmitters requires special consideration in terms of velocity and location tracking.  The second factor is having multiple humans, which converts the single-input communication system into multiple-input ones. This scenario becomes further challenging when the goal is to detect a single type of aerosol (virus or VOC)  emitted from different transmitters. As such, the communication system can suffer from interference that needs distinctive designs to identify the intended transmitter. However, based on the capabilities of existing bio-sensors, detecting different types of aerosol leads to a favorable scenario called ``orthogonal transmission and detection scenario'' as long as there is no reaction between these aerosol streams.

\subsection{Biological Receiver} 
The second component of the system is the receiver which is a bio-compatible machine as depicted in Fig. \ref{block}. The receiver should detect and decode the information sent out by the biological transmitter. The detecting machine is considered one of the most significant components in the proposed system, where the researchers have high designing freedom degree. For instance, this machine can be placed in a room or entry points, or it can be mobile to monitor particular diseases over a wider region. 
The applications can go beyond diseases diagnosis, and the machine could serve as a dedicated device for a particular user (source) that performs various jobs such as health monitoring, mood estimator, etc., where it is part of a network that records and analyzes received data. In addition to detection, we also propose that the receiving machine can play a more \textit{active} role and intelligently respond to decoded messages. For instance, in response to positive virus detection, it can release chemicals or antibodies that limit the spread of that particular virus/disease. The applications that spring out from the development of such machines can be micro or macro-scale. However, the aerosols that the machine would be dealing with are micro-sized, and thus the capabilities of these receivers depend on advancements in nanotechnology or \textit{nanomachines} in particular.

In the context of virus-laden aerosol detection or a reception process in general, MC receivers are particularly relevant \cite{noise2}. As discussed in \cite{aerosol2018}, the reception process can be broken down into three stages:  sampling the breath, sensing/detecting the viral aerosols through biosensors and making an appropriate decision. The commercially available samplers are based on the principle of electrostatic precipitation and can sample nanosized particles in the air with 80-90\% efficiency. Biosensors are devices that sense the presence of specific biological entities and translate them into processable information/signal such as voltage or current in electrical biosensors. An MC receiver based on nanoscale Silicon Nanowire (Si-NW) FET has been discussed in detail in \cite{fet4}. These transistors have antigens placed on Nanowire channel, and binding events between these antigens and virus result in a conductance change across the source-drain channel of the transistor.  The concept of orthogonal detection can be used in identifying different types of viruses using Si-NW FET when several kinds of antibodies are used \cite{fet4}.  We believe that the high selectivity and sensitivity of these MC receivers makes them perfectly suitable for our setup and their existence not only boost the feasibility of our proposed concept but also takes us closer to realizing a fully-functional and robust system. The biosensor is followed by another smart block (controller) which is the brain of our system and is programmed by estimation and decision making algorithms to achieve the end goals.

Similar to the transmitter, both the mobility and the quantity of the receiver units play an important role in determining the machine's capabilities. For the initial study, the receiver can be assumed to be a stationary device at a fixed location. Then, for advanced designs, the receiver can be mobile by attaching it to a drone or robot; then its path can be further controlled and optimized according to different parameters such as the density of crowd/sources. In the case of a network of multiple receiving machines, each unit can sync with other devices and collaborate to perform a joint detection. As a result, the system can have accurate localization, improved coverage, and reliable detection capabilities especially if the  resources are tuned.

\subsection{Transmission Channel}
The transmission channel captures the impact of everything between the biological transmitter and receiver. From communication systems perspective, it is the most critical component since it accounts for all physical characteristics of the communication medium that can \textit{obstruct} the message from reaching the desired destination. The aerosol transmission is  significantly affected by the distance  from the source and the receiver for a specific emitted aerosol volume. Since, the propagation of aerosol is governed by diffusion mechanism, it has limited propagation distance \cite{farsad2016comprehensive}.
One way to extend the travel distance is to employ artificial wind towards the receiver's direction. Different studies show a positive correlation between the presence of wind and distance traveled by virus-laden aerosols \cite{wind1, wind2}. Thus, for particular applications, it is recommended to design artificial wind along with appropriate detection set-up to improve the propagation range/travel distance and guarantee delivering the biological information. The wind velocity plays an essential role in determining the range limitations of the proposed setup where it is responsible for an advection propagation that needs to be incorporated in channel modeling. Since channel modeling involves fluid transport which is itself an extensive, we discuss it separately in the next section.

\section{Aerosol Channel Modeling}\label{sec:channel}
An accurate channel model is essential for not only theoretical analysis but also for the design of an optimum receiver/detector. The model describes the dynamics that are responsible for driving the message from the source to the receiving machine. We begin this section with some basic definitions and explain the basic terminology before outlining the modeling process itself.

 As mentioned earlier, the (artificial) wind acts as the \textit{carrier} that plays a crucial role in transporting aerosols to remote machines. This aerosol transportation is an example of fluid flow where two major processes are responsible for the motion of aerosols known as advection and diffusion. The advection (also known as convection) is due to the wind that can be parameterized by wind velocity. As for the diffusion, it is broken down into two types, molecular diffusion, and turbulent diffusion. The inherent tendency of molecules to reach an equilibrium results in thermal motions known as molecular diffusion. Whereas, the diffusion or mass transfer that results from turbulent eddies is known as turbulent diffusion. The diffusion process is characterized by diffusivity coefficient and the flux changes due to molecular diffusion can be approximated by Fick's law \cite{equation}. It must be noted that from the perspective of aerosol communication proposed in this work, where advection plays a significant role, molecular diffusion is \textit{negligible} compared to turbulent diffusion and is often ignored in the modeling process.
On the contrary, diffusion-based MC focuses on molecular diffusion \textit{only}, and the modeling process is primarily based on Fick's law. These differences in the fluid dynamics result in entirely different models for the two channels. For ease of analysis, it is also assumed that the changes in density in flow field are negligible making the flow \textit{incompressible}.

The mathematical model aims to derive the aerosol concentration along different system stages. To this end, we use the law of mass conservation to describe the system dynamics that result from the introduction of single or multiple aerosol sources into the system.  Sometimes, these aerosols are subjected to elimination from the system by some inactivation process such as the absorption at ground or collection at the receiver side. To analyze the behavior of aerosol motion in both spatial and temporal domains, we use the well-known Navier-Stokes equation that can be combined with the continuity equation to describe the system mathematically. These partial differential equations are subjected to some initial and boundary conditions and are solved to yield an expression for aerosol concentration,  which is not a straightforward process in general \cite{equation, arya1999air, aerosol2018}.  

\subsection{Deterministic Modeling}
Developing deterministic models for the channel involves solving the previously discussed partial differential equation sets (Navier-Stokes and continuity equations) considering boundary and initial conditions. There are two scenarios for the deterministic modeling: steady state and transient analysis.  
As for the former one, having defined a simplified set of boundary conditions and approximating breathing as a constant continuous source at a fixed location, the solution is presented by Gaussian Plume
model \cite{aerosol2018}.  In this model, the concentration profile at a fixed point along the downwind direction (line of sight from source to machine) takes a Gaussian
shape along the centerline. Moreover, as we move away from the source in the downwind
direction, the standard deviation increases. Thus, it is like a set of Gaussian curves (in $y - z$ plane) of increasing variance stacked along $x$-axis as we move away from the source towards the detector direction. As for the transient analysis, we analyze the concentration change due to a single breath, cough or sneezing. The solution of this scenario is presented by Gaussian Puff model \cite{arya1999air} and will be illustrated graphically in Section IV.
\vspace*{-12pt}
\subsection{Stochastic Modeling}

 The inherent randomness of a fluid motion complicates the computation of particles' concentration with certainty. Moreover, the flow-dependent nature of turbulent motion and its non-linear behavior make it extremely difficult to reach a tractable ideal model \cite{taylor1922diffusion, arya1999air, equation}. The most straightforward approach, in this case, is to follow a random walk model, while the most complicated one is to derive a solution to a set of stochastic differential equations. Deriving a close form expression  or the density function of this random process is a challenging problem. Therefore, we resort to following the trajectory of fluid elements at each point in time to simulate the turbulent flow \cite{taylor1922diffusion, arya1999air, equation}.

 The dispersion of particles results from a random velocity component that is composed of drift (mean) and stochastic variations. The simplified Gaussian plume model gives an analytical solution that provides a \textit{mean} concentration profile. However, for accurate simulation of physical systems with fluid flow, stochastic models are required. There are two commonly used computational models based on the fluid flow specifications known as Lagrangian and Eulerian \cite{arya1999air, equation}. The Lagrangian approach focuses on particles directly by tagging and tracking them to monitor the properties of interest such as location, velocity, etc.
On the other hand, the Eulerian approach considers a fixed frame of reference or control volume \textit{across} which the properties of the fluid are tracked. Instead of properties of individual particles, the Eulerian approach tracks the behavior of it's  {markers}as fluid flows across them. Thus, the computation boils down to solving appropriate differential equations for marked locations at each sampling instant.

When dealing with statistical analysis, again the same two approaches are followed. In the Eulerian approach \cite{taylor1922diffusion, arya1999air, equation}, the starting point is a set of instantaneous differential equations which are used to derive other equations governing the known statistics (mean and variance of velocity). The models for unknown quantities based on the known statistics are already provided which are used to reach a set of closed equations for these unknowns. The Lagrangian approach focuses on fluid properties such as velocity, and computes the location of particles based on the statistical description of these properties and conservation equations. 
\begin{figure*}\label{model1}
	\centering
\includegraphics[width=5in]{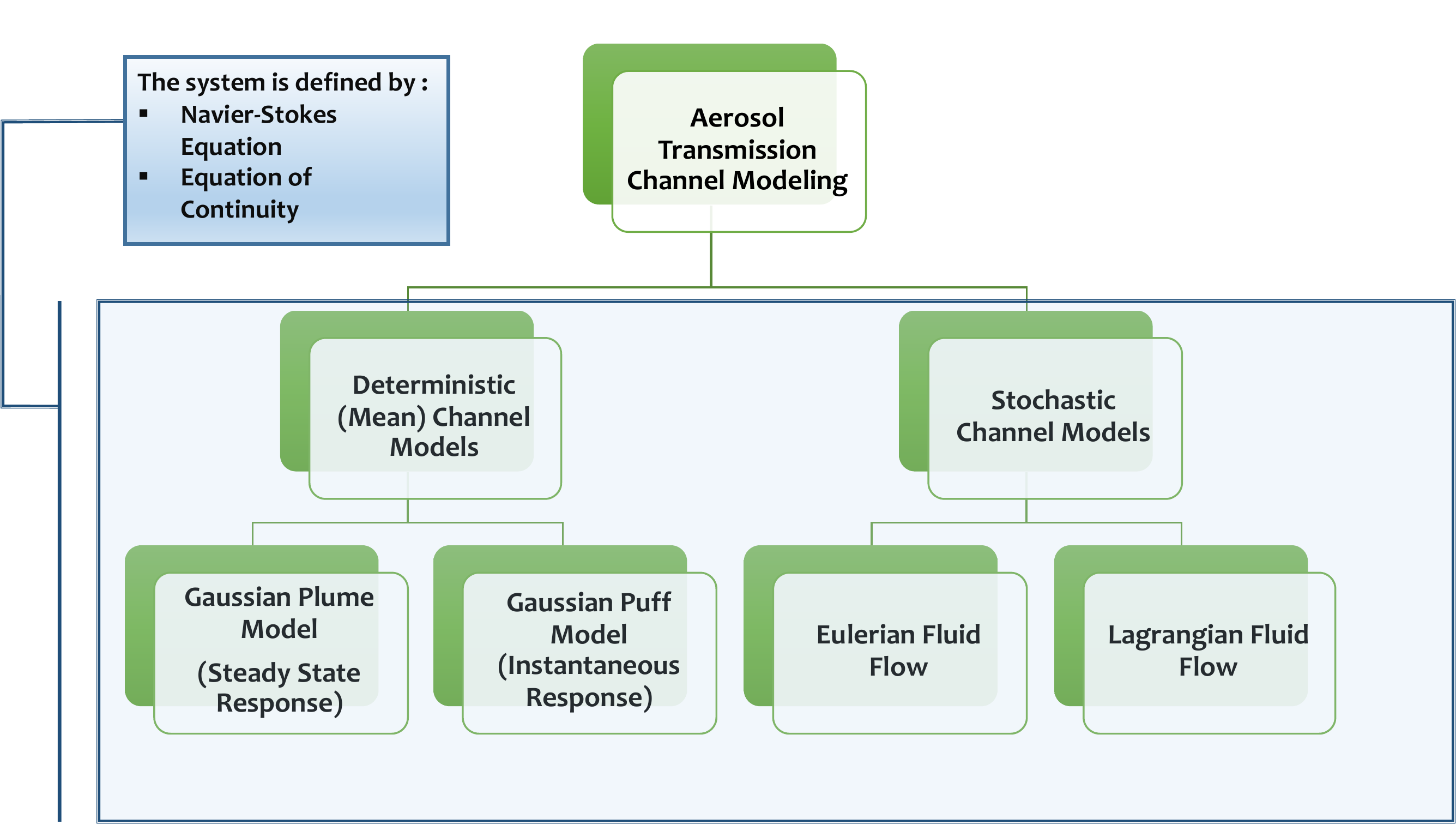}
\caption{Overview of aerosol communication channel modeling.}
\end{figure*}

\section{Viral Aerosol Transmission and Detection}\label{sec:case}

\begin{figure*}
	\centering
\includegraphics[width=6in]{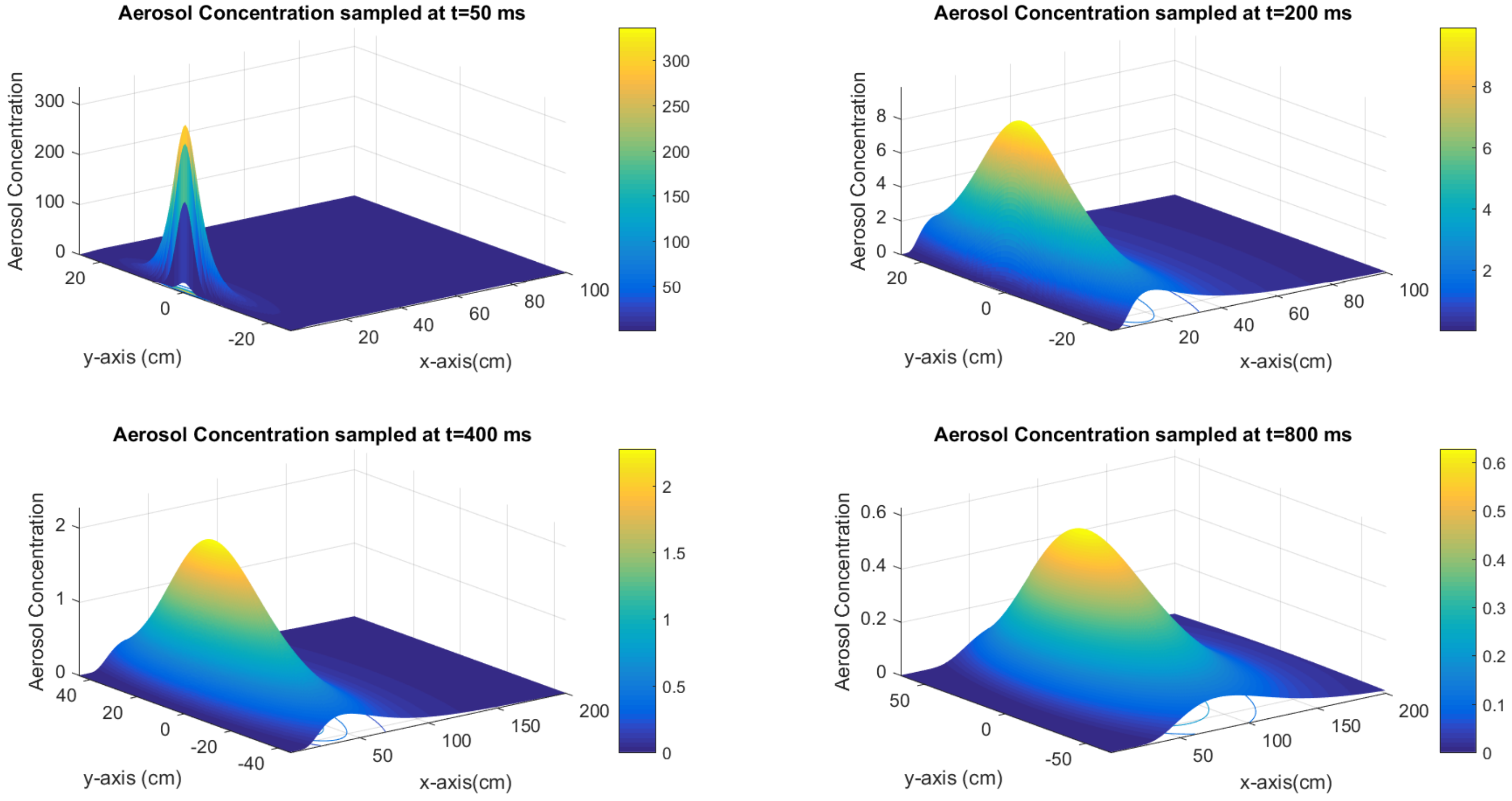} 
\caption{Aerosol concentration sampled at different times in $x-y$ plane along $z=H$ for $Q=40000$, wind speed $1 {\rm{m}}/{\rm{s}}$ and diffusivity coefficient of $0.03 {\rm{m}}^2/{\rm{s}}$.}
\label{inst}
\end{figure*}

Throughout this section, we consider a case study of detecting a virus from the aerosol of infected human breath. To understand the implications of the channel behavior of the proposed communication system, we need to understand the ``symbol'' analogy to the conventional communication system. In this example, the aerosol concentration from a single source over time defines the symbol that carries the useful information. Similar to communication systems, where the impulse response is used to characterize the channel behavior, we consider an impulse source in the spatiotemporal dimension. To this end; we consider an instantaneous transmitter of height $H$ is located at the origin and releases a large number of aerosols $Q$ in the air at time $t=0$. The concentration of aerosols is computed by Gaussian puff model and sampled at 50, 200, 400 and 800 milliseconds (ms) as depicted in Fig. \ref{inst}. It is assumed that major wind component is along $x$-axis (downwind direction) and other components (crosswind direction) are negligible. At $50 ms$, we observe that the aerosol particles are concentrated around the source location, i.e., the origin. Other aerosol samples move forward along the downwind direction and spread over the spatial domain with decreased peak values. Therefore, the breath communication occurs over a dispersive fading channel with long-tail shape, which causes interference between symbols from current sources and the ones that existed in the system before, and latency. In other words, the channel has frequency selective characteristics that can impose intersymbol interference (ISI) on different adopted systems.  

Additionally, we find that the viral aerosol concentration can be still detected for a longer time after the infected human leaves the test room \cite{aerosol2018}.  Unlike electromagnetic signals that travel at a speed of light, the aerosol droplets have very slow propagation speeds with a lag of seconds, which requires special design consideration. The proposed system can suffer from several types of challenges that should be effectively treated before system realization.

\subsection{Research and Development Challenges}
At this point, we highlight the main challenges that need to be addressed in designing breath-based human-to machine communication systems. 
 
\subsubsection{Dynamic Nature of Biological Transmitter} 
Compared to the conventional system, the proposed breath communication system has different transmission features.  The unexpected body activity grants dynamic transmitter characteristics that affect the communication link quality. Although the breathing process is a continuous regular process, the exhaled breath might not be the same at all times even for the same human due to different activities. Thus, both the ISI and sampling time change and cannot be easily determined.  Besides, it is also possible that the \textit{desired markers} ratio in the exhaled breath varies so much and leads to false results. 
 
\subsubsection{Detecting Multiple Sources}
The existence of multiple sources that release the same or different types of viral aerosol presents a challenge on the receiver side. Firstly, assume that two sources release the same kind of viral aerosols. Achieving accurate detection requires developing complex detection techniques to mitigate the possible interference between these sources. Secondly, if the sources release different types of viral aerosols, it is important to study possible interaction between them in the atmosphere before they reach the receiver side. Therefore, to design a robust receiver, it is compulsory to study the dynamics of transmitting entities and model both the variations and interactions. This challenging task requires not only probing into literature from fluid dynamics but also biology. However, if there is no interaction between the viral aerosols, then the detection will not be affected thanks to the orthogonal transmission.
 
\subsubsection{Transmission Limitations}
Controlling the transmitter flow rate, encoding, message duration, and emission frequency is severely limited in the proposed communication system compared with the conventional one.  We find that the receiver might be \textit{blind} to the biological transmitter in most of the applications. These restraints on transmitting data control and knowledge of transmitted sequence put an additional burden on the receiving device and intensify the design process. 
 
\subsubsection{Analytical Channel Models}
The mathematical model for the viral aerosol transmission and detection is obtained by solving a set of partial differential equations under some assumptions and boundary conditions that simplify the analysis as mentioned in Section \ref{sec:channel}. The derivation process can change dramatically if either the assumptions or conditions is relaxed. For example, the walls can either absorb or reflect the aerosol with varying degrees, which can change the derived model entirely.  Moreover, closed form solutions may not be possible, which complicates system analysis and design. Moreover, the estimation of system parameters, such as the diffusivity coefficient, is another big challenge.
 
\subsubsection{Synchronization}  
Adjusting the time alignment to a specific reference is necessary to establish reliable communication links. The machine should know \textit{when} the transmitter started talking to decode the received signal correctly. Synchronization is accomplished by a master clock or a training-based approach as in conventional communication systems. However, the situation is not clear for this proposed setup, and therefore, the design of an appropriate synchronization mechanism is itself a big challenge.

\section{Open Issues} \label{sec:issues}
The newly proposed concept in this article is still in its early research stages with some promising results \cite{aerosol2018}. There are several grey areas that need clarifications and numerous open issues that require investigations to provide  solid foundations before the development phase.
 
\begin{figure}[h!]
\centering
\includegraphics[width=3in]{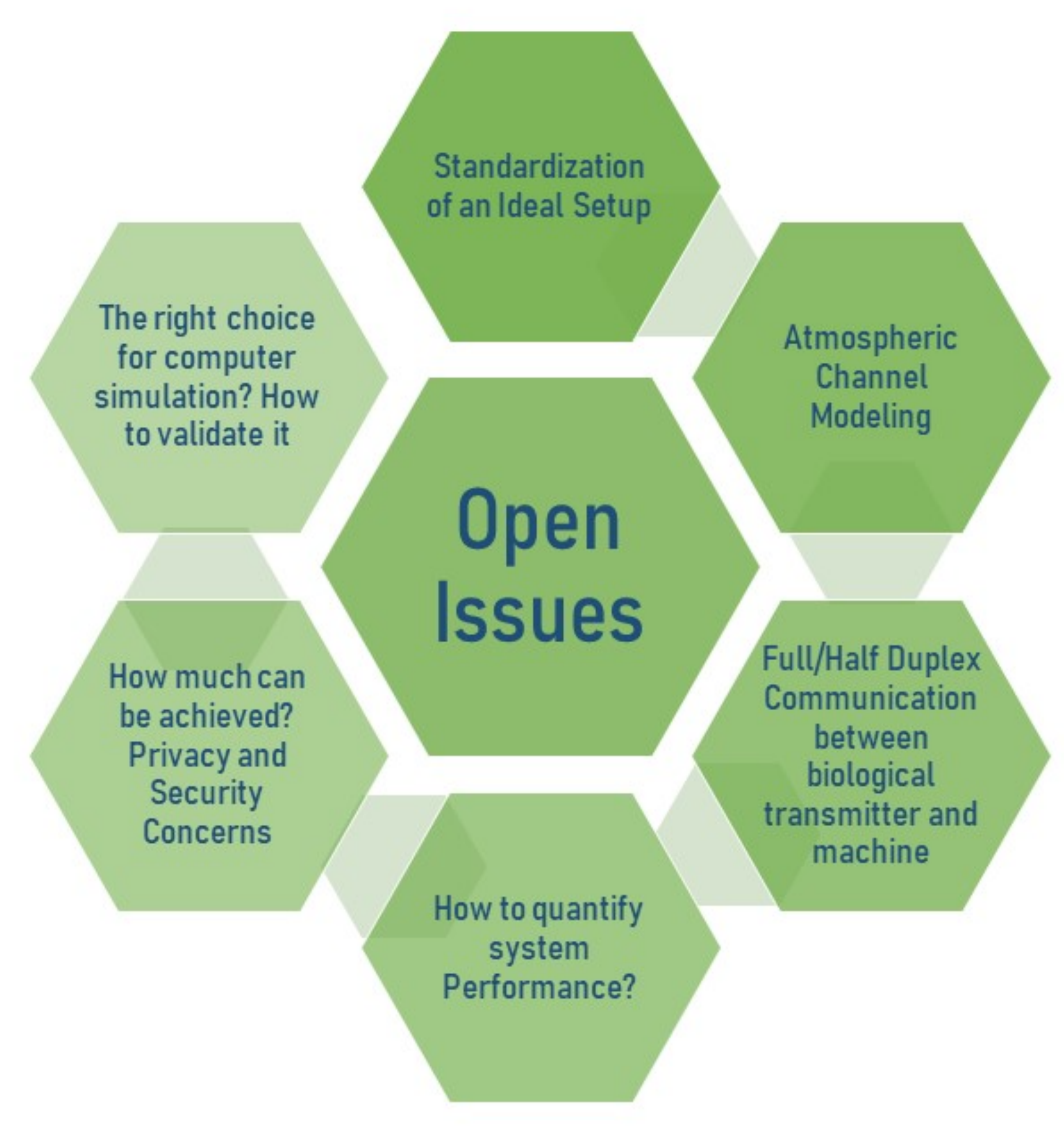}
\caption{Summary of open issues in research and development.}
\end{figure}

\subsection{How ``ideal'' an ideal system should be?}
It is a common practice in science to assume some ideal conditions that simplify the analysis. As mentioned in \ref{sec:channel}, the commonly deployed Gaussian plume or puff channel models are derived under simplifying assumptions such as incompressible flow, boundaryless environment, and flat ground. However, the practicality of these simplifications is still not clear and needs a debate. Thus, it is essential to develop flexible models that maintain a balance between analytical complexity and  closeness to the real-life scenario.  

\subsection{Choices of stochastic channel model} 
Stochastic channel modeling is described from the perspective of atmospheric dispersion in Section \ref{sec:channel}. In this context, we can deploy computational fluid dynamics using the Eulerian or Lagrangian method to simulate aerosol dispersion. Each method has its benefits and limitations that are not stated due to space constraints. However, the choice between the two models is still an open question. Moreover, modeling of external and internal noise at machine is also an unexplored area. 

\subsection{Two-way communication}
The receiving machine can act as an active device and respond in a particular manner after decoding the message embedded in exhaled breath  as discussed in Section \ref{sec:block}. This concept paves the way for two-way communication between the machine and biological transmitter through \textit{half} or \textit{full duplex} systems. Some of the crucial challenges associated with realization of such systems are interference mitigation and latency minimization. To deal with these problems, we need to consider interdisciplinary research between ICT, fluid dynamics and biological fields. For example, the machine can release a predesigned VOC that triggers the inner human body to release another/same VOC as a response to the transmitted message.

\subsection{Defining performance metrics}
Performance metrics are used to characterize and measure system functionality in delivering required services. For instance, the throughput is used to quantify the communication channel ability in transmitting high data rates, whereas energy efficiency is adopted as an energy-aware metric to monitor energy consumption and harvesting along transceivers. In the proposed system, we need to choose or define metrics that can accommodate several applications envisioned from the concept of breath communication. For instance, in virus diagnosis application, we can use both miss-detection and false-alarm probabilities to measure the system performance.

\subsection{Coexistence with existing technologies}  
As a vital participant in the HBC, the breath communication systems need to be integrated with other systems in the internet-of-everything (IoE) network. Exchanging information between IoE network and the proposed biological system is challenging and requires in-depth investigations. The proposed integration can help in providing a lot of remote access services in several fields like medicine and environmental science. 

\subsection{Simulation and validation}
Before the development of a real-life breath-based communication system, a light-weight, flexible, and powerful simulation software is required. There are several open-source software, such as ANSYS, which are available for fluid flow simulations. However, the choice of the appropriate software model that accurately describes the proposed system is not clear. Besides, there is a need to develop mechanisms to validate the models and simulation setup.

\subsection{Privacy and security}
 The direct involvement of humans in the proposed communication system raises security concerns on the personal data. The receiver is vulnerable to attacks from malignant machines that not only put user data at risk but also pose threats to system functionality. Therefore, there is a need to design a biocompatible lightweight and effective security system that can provide an authorization framework to secure personal data privacy.

\section{Conclusion}\label{sec:conc}
We have proposed communication through breath as a new enabler paradigm for the HBC. The human breath acts an information source that reports biological characteristics through exhalation and communicates with human body through inhalation. We have analyzed the proposed concept as a communication problem and explored the system architecture. Furthermore, we have highlighted several challenges and open research issues concerning communication perspective. However, to master the research in this direction, interdisciplinary research should be conducted to integrate concepts from ICT, fluid dynamics and biology. Thus, we expect to have interactive communications between human and machines that can encode and decode human breath resulting in a wide range of innovative applications. 

\bibliographystyle{IEEEtran}
\bibliography{IEEEabrv,references}

\begin{thebibliography}{10}
\providecommand{\url}[1]{#1}
\csname url@samestyle\endcsname
\providecommand{\newblock}{\relax}
\providecommand{\bibinfo}[2]{#2}
\providecommand{\BIBentrySTDinterwordspacing}{\spaceskip=0pt\relax}
\providecommand{\BIBentryALTinterwordstretchfactor}{4}
\providecommand{\BIBentryALTinterwordspacing}{\spaceskip=\fontdimen2\font plus
\BIBentryALTinterwordstretchfactor\fontdimen3\font minus
  \fontdimen4\font\relax}
\providecommand{\BIBforeignlanguage}[2]{{%
\expandafter\ifx\csname l@#1\endcsname\relax
\typeout{** WARNING: IEEEtran.bst: No hyphenation pattern has been}%
\typeout{** loaded for the language `#1'. Using the pattern for}%
\typeout{** the default language instead.}%
\else
\language=\csname l@#1\endcsname
\fi
#2}}
\providecommand{\BIBdecl}{\relax}
\BIBdecl

\bibitem{dixit2017human}
S.~Dixit and R.~Prasad, \emph{Human Bond Communication: The Holy Grail of
  Holistic Communication and Immersive Experience}.\hskip 1em plus 0.5em minus
  0.4em\relax John Wiley \& Sons, 2017.

\bibitem{almstrand2011analysis}
A.-C. Almstrand, ``Analysis of endogenous particles in exhaled air,'' PhD
  thesis, Institute of Medicine at Sahlgrenska Academy, University of
  Gothenburg, Gothenburg, Sweden,, 2011.

\bibitem{de2014review}
B.~de~Lacy~Costello, A.~Amann, H.~Al-Kateb, C.~Flynn, W.~Filipiak, T.~Khalid,
  D.~Osborne, and N.~M. Ratcliffe, ``A review of the volatiles from the healthy
  human body,'' \emph{J. Breath Research}, vol.~8, no.~1, p. 014001, 2014.

\bibitem{flu1}
P.~Fabian, J.~J. McDevitt, W.~H. DeHaan, R.~O. Fung, B.~J. Cowling, K.~H. Chan,
  G.~M. Leung, and D.~K. Milton, ``Influenza virus in human exhaled breath: an
  observational study,'' \emph{PloS one}, vol.~3, no.~7, p. e2691, 2008.

\bibitem{foot}
Z.~Xu, F.~Shen, X.~Li, Y.~Wu, Q.~Chen, X.~Jie, and M.~Yao, ``Molecular and
  microscopic analysis of bacteria and viruses in exhaled breath collected
  using a simple impaction and condensing method,'' \emph{PLOS ONE}, vol.~7,
  no.~7, pp. 1--8, 07 2012.

\bibitem{voc}
W.~Miekisch, J.~K. Schubert, and G.~N{\"o}ldge-Schomburg, ``Diagnostic
  potential of breath analysis--focus on volatile organic compounds.''
  \emph{Clinica Chimica acta; Intern. J. Clinical Chemistry}, vol. 347 1-2, pp.
  25--39, 2004.

\bibitem{farsad2016comprehensive}
N.~Farsad, H.~B. Yilmaz, A.~Eckford, C.-B. Chae, and W.~Guo, ``A comprehensive
  survey of recent advancements in molecular communication,'' \emph{{IEEE}
  Commun. Surveys Tuts.}, vol.~18, no.~3, pp. 1887--1919, 2016.

\bibitem{aerosol2018}
M.~Khalid, O.~Amin, S.~Ahmed, and M.-S. Alouini, ``System modeling of virus
  transmission and detection in molecular communication channels,'' in
  \emph{IEEE Intern. Conf. Commun. (ICC18)}.\hskip 1em plus 0.5em minus
  0.4em\relax Kansas, USA: IEEE, 2018, pp. 1--6.

\bibitem{noise2}
M.~Kuscu and O.~B. Akan, ``On the physical design of molecular communication
  receiver based on nanoscale biosensors,'' \emph{IEEE Sensors J.}, vol.~16,
  no.~8, pp. 2228--2243, 2016.

\bibitem{fet4}
F.~Shen, M.~Tan, Z.~Wang, M.~Yao, Z.~Xu, Y.~Wu, J.~Wang, X.~Guo, and T.~Zhu,
  ``Integrating silicon nanowire field effect transistor, microfluidics and air
  sampling techniques for real-time monitoring biological aerosols,''
  \emph{Environmental science \& technology}, vol.~45, no.~17, pp. 7473--7480,
  2011.

\bibitem{wind1}
X.~Wu, Y.~Lu, S.~Zhou, L.~Chen, and B.~Xu, ``Impact of climate change on human
  infectious diseases: Empirical evidence and human adaptation,''
  \emph{Environment International}, vol.~86, pp. 14 -- 23, 2016.

\bibitem{wind2}
P.-S. Chen, F.~T. Tsai, C.~K. Lin, C.-Y. Yang, C.-C. Chan, C.-Y. Young, and
  C.-H. Lee, ``Ambient influenza and avian influenza virus during dust storm
  days and background days,'' \emph{Environmental health perspectives}, vol.
  118, no.~9, pp. 1211--1216, Sep. 2010.

\bibitem{equation}
S.~B. Pope, \emph{Turbulent Flows}.\hskip 1em plus 0.5em minus 0.4em\relax
  Cambridge University Press, 2000.

\bibitem{arya1999air}
S.~P. Arya, \emph{Air pollution meteorology and dispersion}.\hskip 1em plus
  0.5em minus 0.4em\relax Oxford University Press New York, 1999, vol. 310.

\bibitem{taylor1922diffusion}
G.~I. Taylor, ``Diffusion by continuous movements,'' \emph{Proceedings of the
  london mathematical society}, vol.~2, no.~1, pp. 196--212, 1922.

\end{thebibliography}

\begin{IEEEbiographynophoto}
 {Maryam Khalid} received the BSc. Degree from University of Engineering and Technology, Lahore, Pakistan in 2015 and the MS degree from Lahore University of Management Sciences (LUMS), Pakistan in 2017. Currently, she is a PhD Student at Electrical and Computer Engineering Department at Rice University, Houston, USA. She is a recipient of John Clark, Jr. Fellowship Award, K2I Computational Science and Engineering Fellowship, LUMS Merit award and several other National level distinctions and awards during her studies in Pakistan. She received a Gold Medal and was placed on Dean?s Honor list in her MS at LUMS. Her research interests include Communication systems, Signal Processing and their applications in bio-inspired domains.
\end{IEEEbiographynophoto}

\vspace*{10pt}
\begin{IEEEbiographynophoto}
 {Osama Amin} (S'07, M'11, SM'15) received B.Sc. degree in Electrical and Electronic Engineering from Aswan University, Aswan, Egypt, in 2000, M.Sc. degree in Electrical and Electronic Engineering from Assiut University, Assiut, Egypt in 2004 and Ph.D. degree in Electrical and Computer Engineering, University of Waterloo, Canada in 2010. In June 2012, he joined Assiut University as an Assistant Professor in the Electrical and Electronics Engineering department. Currently, he is a research scientist King Abdullah University of Science and Technology (KAUST), Thuwal, Makkah, Kingdom of Saudi Arabia. His general research interests lie in communication systems and signal processing for communications with special emphasis on wireless applications.
\end{IEEEbiographynophoto}

\begin{IEEEbiographynophoto}
 {Sajid Ahmed} completed his PhD in Digital Signal Processing at the King's College London and Cardiff University, UK in 2005. Presently, he is a faculty member in the Information Technology University (ITU) Lahore, Pakistan. Before, Joining ITU, he was a researcher at the Queen's University Belfast and the University of Edinburgh, and research scientist at the King Abdullah University of Science and Technology (KAUST), Thuwal, Kingdom of Saudi. Dr. Ahmed's current research interests include the linear and non-linear optimization techniques, low complexity parameter estimation for communication and radar systems, and waveforms design for MIMO radar. He is a Senior member of IEEE.
\end{IEEEbiographynophoto}

\begin{IEEEbiographynophoto}
 {Basem Shihada} is an associate and founding professor of computer science and electrical engineering in the Computer, Electrical and Mathematical Sciences $\&$ Engineering (CEMSE) Division at King Abdullah University of Science and Technology (KAUST). Before joining KAUST in 2009, he was a visiting faculty at the Computer Science Department in Stanford University. His current research covers a range of topics in energy and resource allocation in wired and wireless communication networks, including wireless mesh, wireless sensor, multimedia, and optical networks. He is also interested in SDNs, IoT, and cloud computing. In 2012, he was elevated to the rank of Senior Member of IEEE.
\end{IEEEbiographynophoto} 

\begin{IEEEbiographynophoto}{Mohamed-Slim Alouini}  (S'94-M'98-SM'03-F'09)  was born in Tunis, Tunisia. He received the Ph.D. degree in Electrical Engineering from the California Institute of Technology (Caltech), Pasadena, CA, USA, in 1998. He served as a faculty member in the University of Minnesota, Minneapolis, MN, USA, then in the Texas A$\&$M University at Qatar,
Education City, Doha, Qatar before joining King Abdullah University of Science and Technology (KAUST), Thuwal, Makkah Province, Saudi Arabia as a Professor of Electrical Engineering in 2009. His current research interests include the modeling, design, and performance analysis of wireless communication systems.
\end{IEEEbiographynophoto}

\vfill
\end{document}